\documentclass[preprint2]{aastex}
\usepackage{natbib}
\usepackage{multicol}
\usepackage{verbatim}
\usepackage{color}
\usepackage{bm}
\usepackage{lscape}
\title{Characterizing Transiting Planet Atmospheres through 2025}
\author{N.B.~Cowan$^1$, T.~Greene$^2$, D.~Angerhausen$^3$, N.E.~Batalha$^4$, M.~Clampin$^3$, K.~Col\'on$^5$, I.J.M.~Crossfield$^6$, J.J.~Fortney$^7$, B.S.~Gaudi$^8$, J.~Harrington$^9$, N.~Iro$^{10}$, C.F.~Lillie$^{11}$, J.L.~Linsky$^{12}$, M.~Lopez-Morales$^{13}$, A.M.~Mandell$^2$, and K.B.~Stevenson$^{14}$, \emph{on behalf of} ExoPAG SAG-10}
\affil{$^1$Amherst College (ncowan@amherst.edu)}
\affil{$^2$NASA Ames Research Center}
\affil{$^3$NASA Goddard Space Flight Center}
\affil{$^4$Pennsylvania State University}
\affil{$^5$Lehigh University}
\affil{$^6$University of Arizona}
\affil{$^7$University of California, Santa Cruz}
\affil{$^8$The Ohio State University}
\affil{$^9$University of Central Florida}
\affil{$^{10}$Universit\"at Hamburg}
\affil{$^{11}$Lillie Consulting}
\affil{$^{12}$University of Colorado}
\affil{$^{13}$Harvard-Smithsonian Center for Astrophysics} 
\affil{$^{14}$University of Chicago}

\begin{document}
\maketitle
\newpage
\section*{Executive Summary}
The discovery of planets around other stars is revolutionizing our
notions of planet formation and is poised to do the same for planetary
climate.  Studying \emph{transiting} planets is complementary to eventual studies of directly-imaged planets: 1) we can readily measure the mass and radius of transiting planets, linking atmospheric properties to bulk composition and formation, 2) many transiting planets are strongly irradiated and exhibit novel atmospheric physics, and 3) the most common temperate terrestrial planets orbit close to red dwarf stars and are difficult to image directly. We have only been able to comprehensively
characterize the atmospheres of a handful of transiting planets,
because most orbit faint stars. The Transiting Exoplanet Survey Satellite (TESS) will discover
transiting planets orbiting the brightest stars, enabling, in
principle, an atmospheric survey of $10^{2}$--$10^{3}$
bright hot Jupiters and warm sub-Neptunes.  Uniform
observations of such a statistically significant sample would provide
leverage to understand---and learn from---the diversity of short-period
planets, and would identify the minority of truly special planets
worthy of more intensive follow-up.  
We argue that the best way to maximize the scientific returns of TESS is
to adopt a triage approach.  A space mission consisting of a $\sim$1 m
telescope with an optical--NIR spectrograph could
measure molecular absorption for non-terrestrial planets discovered by TESS,
as well as eclipses and phase variations for the hottest jovians. Such a mission could observe up to $10^{3}$
transits per year, thus enabling it to survey a large
fraction of the bright ($J<11$) hot-Jupiters and warm sub-Neptunes TESS
is expected to find. The James Webb Space Telescope (JWST) could be used to perform detailed atmospheric characterization of the most interesting transiting targets (transit, eclipse, and---when possible---phase-resolved spectroscopy).  TESS is also expected to discover
a few temperate terrestrial planets transiting nearby M-Dwarfs.
Characterizing these worlds will be time-intensive: JWST will need months to provide tantalizing constraints on the presence of an atmosphere, planetary rotational state, clouds, and greenhouse gases.  Future flagship missions should be designed to provide better constraints on the habitability of M-Dwarf temperate terrestrial planets.

\newpage
\section{Context}
The study of exoplanet atmospheres has exploded in the past decade.  In 2013, the Exoplanet Exploration Analysis Group (ExoPAG) created---with approval from NASA's Astrophysics Subcommittee---a tenth Study Analysis Group (SAG-X) to consider what NASA could do in the next decade to better understand the atmospheres of transiting planets.  SAG-X had open membership and involved three presentations to the exoplanet community. The first presentation, outlining the challenges and opportunities of studying transiting exoplanet atmospheres, was made at the ExoPAG~8 meeting preceding the Oct.\ 2013 Division of Planetary Sciences meeting in Denver, CO.  We held a mini-workshop on the capabilities of the James Webb Space Telescope for characterizing transiting exoplanets at ExoPAG~9 preceding the Jan.\ 2014 American Astronomical Society meeting in Washington, DC.  Interested members of the community helped draft the current document over the course of the 2014 calendar year and we presented it at ExoPAG 11, preceding the Jan.\ 2015 AAS meeting in Seattle, WA.  Multiple drafts of this report have been circulated to the ExoPAG membership and the exoplanet community at large and we have done our best to implement feedback. This document is therefore a consensus view of what can and should be done in the field of transiting exoplanet atmospheres in the next decade.

\section{Planetary Science from the Top--Down}
\textbf{Our knowledge of Earth and the Solar System planets will always exceed our knowledge of any \emph{individual} exoplanet, but the diversity of exoplanets enables the \emph{statistical} study of planets to crack difficult problems in planetary science.}  

What started as a trickle in the mid 1990's is now a torrent, with over one thousand extrasolar planets currently known, and thousands of candidates awaiting confirmation. The study of exoplanets has already revolutionized our view of planet formation, and will soon do the same to our understanding of planetary atmospheres and interiors. The diversity of exoplanets gives us the leverage to crack hard problems in planetary science: cloud formation, atmospheric circulation, plate tectonics, etc. However, the characterization of exoplanets presents a challenge familiar to astronomers: our targets are so distant that we only see them as unresolved dots.

Many aspects of planetary science are currently accessible for exoplanets, or soon will be. Since we observe exoplanetary systems from the outside, the easiest aspect to constrain is the architecture of planetary systems, and indeed our theories of planet formation are currently being revolutionized by our growing knowledge of planetary demographics and architecture. Transiting planets are crucial to our understanding of planet formation, because they are the only planets for which we can hope to know the orbital architecture, bulk density, \emph{and} atmospheric composition.

The study of individual planets is likewise progressing from the top--down: first the exospheres, then the atmospheres, and the surfaces last. It is therefore difficult to make definitive statements about the surface conditions of exoplanets, and their interiors will only be known to us through their bulk density, and surface character.  The study of planetary atmospheres, however, is poised to be revolutionized by observations of ``exoclimes.''  The first such measurements constrained planetary-scale temperature structure and composition, while the second generation of measurements leveraged the orbital motion of the planet to infer the horizontal temperature structure of the planet.  Planned instruments will enable 3D measurements of atmospheric composition, temperature structure, and winds.
 
\section{Planetary Climate}
\textbf{A planet's albedo determines how much radiation it absorbs, atmospheric composition dictates how energy trickles up via radiation and convection, while atmospheric (and oceanic) dynamics determine how heat is transported from regions receiving more sunlight to those receiving less.}

\begin{figure}[!htb]
\begin{center}
\vspace{-0.1cm}
 \includegraphics[width=40mm, angle=90]{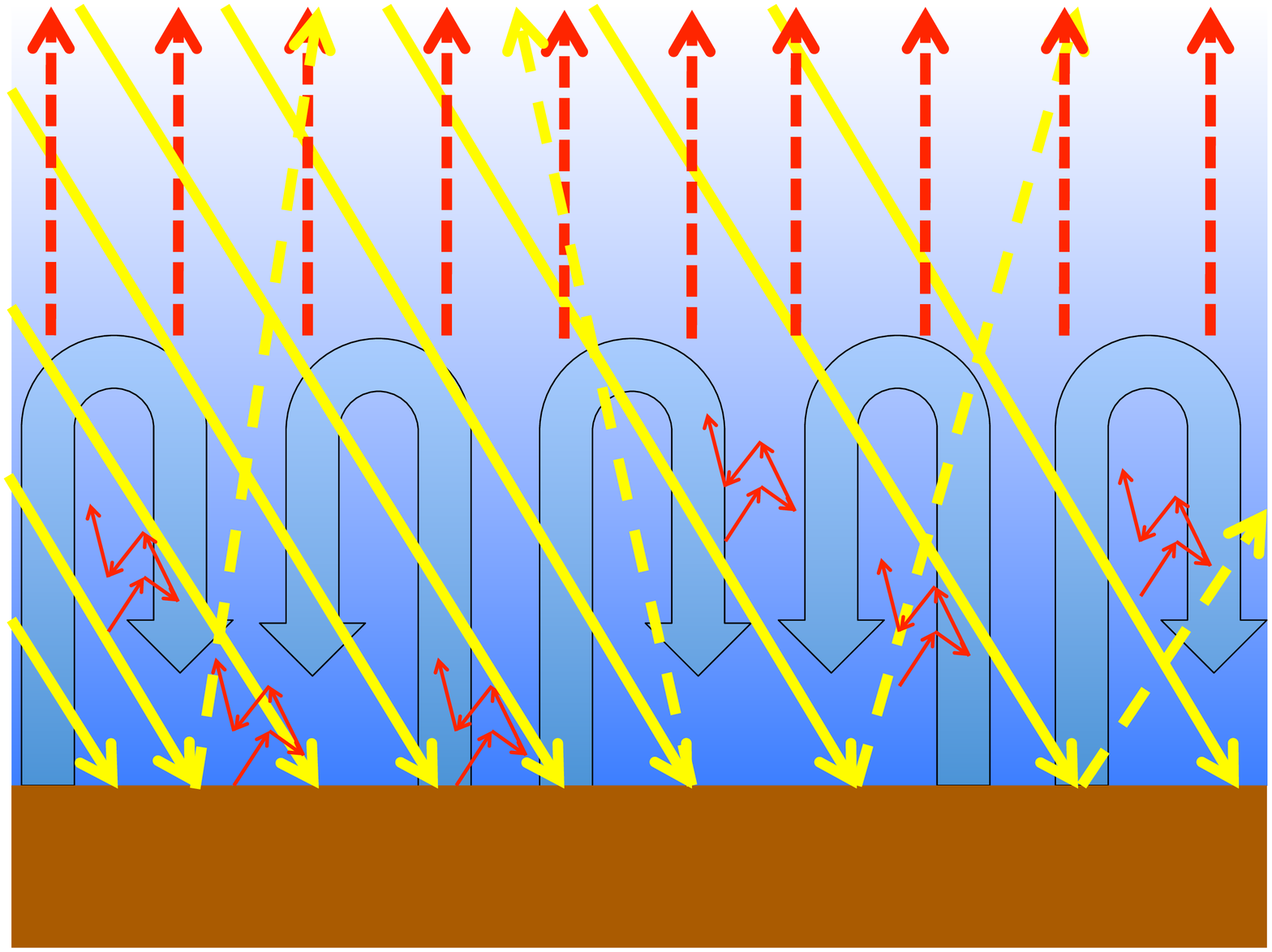}
    \caption{Planetary climate is determined by stellar radiation interacting with a planetary atmosphere.  Incoming radiation (solid yellow lines) is either reflected (dashed yellow lines) or absorbed.  The warm air emits at longer wavelengths than the incoming light. Longwave radiation is readily absorbed by greenhouse molecules in the atmosphere, resulting in radiative diffusion (small red arrows). The inefficient upward heat transport means that the lower parts of the atmosphere tend to be hotter than the overlying regions, and in practice most planetary atmospheres convect over some pressure range (blue arrows). The cooler, thinner upper regions of the atmosphere emit thermal radiation (dashed red lines) that balances the absorbed shortwave radiation.         \label{climate_cartoon}}
\vspace{-0.1cm}
\end{center}
\end{figure}

The primary aspect of planetary climate is temperature, averaged over time, and often space. A detailed calculation of climate involves radiative transfer, fluid dynamics, chemistry and, in the case of an inhabited planet, biology. Stripped to its essentials, however, climate describes the interaction of star light with a planetary atmosphere (Figure~\ref{climate_cartoon}).  

\begin{figure*}[!htb]
\begin{center}
\vspace{-0.1cm}
 \includegraphics[width=50mm, angle=-90]{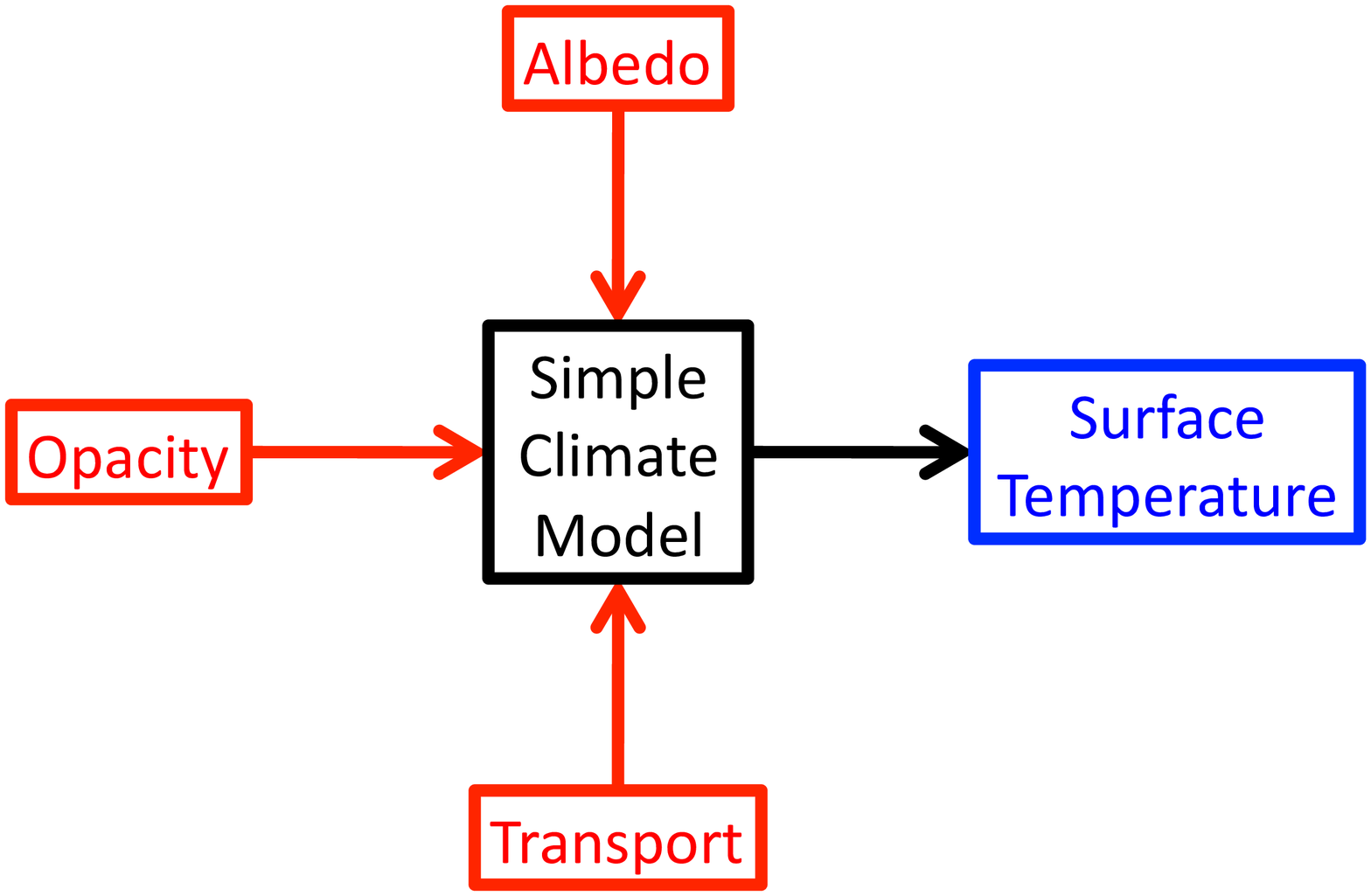} \includegraphics[width=50mm, angle=-90]{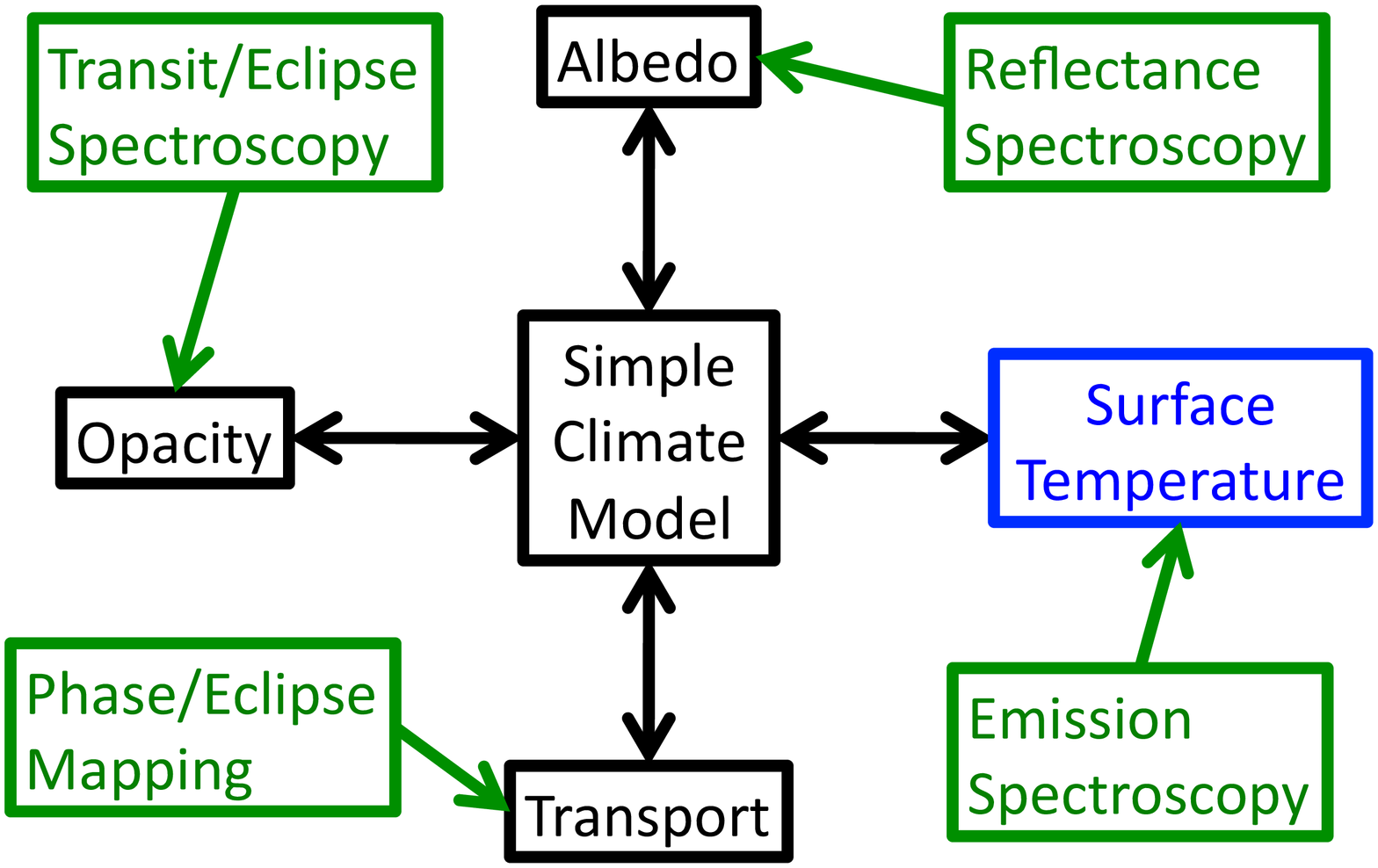}
    \caption{\emph{Left:} It is possible to predict a planet's climate given its albedo (reflectiveness), atmospheric opacity (greenhouse gas abundances), and heat transport (wind speeds and ocean circulation). Unfortunately, these critical inputs cannot be predicted in general. \emph{Right:} Fortunately, it is possible to measure planetary albedo, atmospheric composition, and heat transport, even for exoplanets (in many cases, there are multiple independent means of constraining atmospheric properties; we list only a few in the interest of clarity).  This makes it possible to empirically determine a planet's climate, even in the absence of a fully predictive theory of planetary climate.  In the long run, such observations may reveal some of the underlying principles governing cloud formation, volatile cycling, and large-scale circulation.           \label{climate_flow}}
\vspace{-0.1cm}
\end{center}
\end{figure*}

A simple energy balance model, shown schematically in the left panel of Figure~\ref{climate_flow}, can therefore be used to predict planetary climate. Unfortunately, despite over a century of research into Earth's climate, there are currently no comprehensive, predictive theories for cloud formation (planetary albedo), volatile cycling (greenhouse gas abundances), or wind speeds (heat transport). The empirical approach to this challenge is to acquire observations for many different planets in the hopes of uncovering the principles of climate (right panel of Figure~\ref{climate_flow}); the first step in that direction has been the study of Solar System worlds over the last half-century.  

The logical next step, already underway, is to characterize the atmospheres of extrasolar planets.  Exoplanets are more diverse than the planets orbiting our Sun, and hence provide more leverage for testing theories.  On the other hand, many exoplanets are similar to those in our Solar System, providing crucial Rosetta stones: temperate terrestrial planets that rotate slowly allow us to empirically verify the effects of Coriolis forces on atmospheric circulation, super-Jupiters let us test the effects of surface gravity on Jovian atmospheres, etc.

While the most detailed atmospheric studies will always focus on Earth, and \emph{in situ} measurements will be limited to the Solar System, the vast majority of planets are extrasolar.  This means that the most extreme worlds, and those most like the Earth, are exoplanets. Only by studying these planets can we hope to develop comprehensive theories of climate. To paraphrase Kipling (1891): \emph{What should they know of Earth who only Earth know?}

\section{Transiting Planets}
\textbf{Transiting planets are representative of planets on short-period orbits. Exoplanets that transit Sun-like stars tend to be hot, while those transiting low-mass stars are merely warm, and represent the majority of temperate terrestrial planets. Transit spectroscopy probes atmospheric opacity, while measurements of thermal emission constrain vertical and horizontal temperature structure. High cadence, high-precision emission spectroscopy throughout a planet's orbit enables 3D mapping of its atmosphere.}

A transiting planet is a planet that passes in between its host star and the observer.  Any planet that orbits a star may be a transiting planet, but transiting planets are essentially synonymous with short-period planets, since the probability that a planet transits is the ratio of the stellar radius to the planet--star separation.  This makes the transit method an improbable way to study Solar System analogs.

Direct imaging is a promising approach to studying the atmospheres of Solar System analogs, but close-in planets are difficult to study in this way, because any means of blocking/nulling the starlight is liable to block the planet, too.  We instead study the combined light of the planetary system, so the dominant source of photons (and photon noise) is the star rather than the planet. The signal-to-noise ratio for measurements of planetary light is therefore proportional to the planetary flux, rather than its square-root; the usual biases in favor of bigger and brighter targets are especially true for transiting planets. 

Characterizing transiting planets would be of merely theoretical interest, except that many exoplanets orbit much closer to their stars than in our Solar System. Some of these unfamiliar worlds are actually rare, like hot Jupiters \citep[][]{Wright_2012}, while others represent a common outcome of planet formation, as with packed systems of warm sub-Neptunes \citep[][]{lissauer2011closely}.

A hot Jupiter is a jovian planet that orbits its host star with a period of $\lesssim$1~week.\footnote{Hot Jupiters should not be confused with \emph{young Jupiters}, the typical targets of current direct-imaging surveys, which are also hot.}  Such planets only exist around $\sim1$\% of Sun-like stars \citep[][]{Wright_2012}, but have a transit probability of 10\%, so there is a transiting hot Jupiter for every few thousand FGK stars. These extreme worlds experience strong radiative forcing resulting in day--night temperature contrasts of hundreds to thousands of K \citep{Showman_2002}.  The high atmospheric temperatures thermally ionize alkali metals, effectively coupling atmospheric dynamics to planetary magnetic fields \citep{Perna_2010a, Perna_2010b}.  The radiative environment of hot Jupiters often results in mass loss, and some are thought to be undergoing Roche lobe overflow onto their host star \citep{Li_2010}.

The most common currently detectable transiting planets in the Galaxy are sub-Neptunes in tight (10--100 days) orbits around their host stars \citep{Howard_2012, Fressin_2013}. These warm sub-Neptunes have masses dominated by rock and ice, but covered in thick H+He atmospheres, and are sufficiently cool (500--1000~K) that photochemistry often trumps thermochemical equilibrium \citep{Moses_2014}.   The \emph{a priori} transit probability of such planets is low, but they are intrinsically common, occurring around roughly half of stars. As such, they form the bulk of the \emph{Kepler} crop.  The densities of these planets suggest they are largely made of ice, and/or have substantial H+He atmospheres, hence their name.  Our experience from the Solar System suggests that these planets likely have high atmospheric metallicities \citep{Fortney_2013}. In short, these planets are expected to be less extreme than hot Jupiters in terms of radiative forcing, but probably more interesting in terms of chemistry.  

Temperate terrestrial planets transiting M-dwarf stars are often touted as the poor-astronomer's Earth analog, since they are easier to detect and characterize than a true Earth twin.  Based on what we currently know, however, M-Dwarf planets are \emph{the most common habitable worlds}.  That is because: 1.\ rocky planets are much more common in the temperate zones of M-Dwarfs \citep{Dressing_2013, Morton_2013} than in the temperate zones of Sun-like stars \citep{Petigura_2013, Foreman-Mackey_2014, Farr_2015}, 2. small stars are more common than big stars \citep[e.g.,][]{Bochanski_2010},  3.\ the tidally-locked nature of these planets is not a challenge to climate \citep{Joshi_1997, Merlis_2010, Edson_2011} and may double the width of the habitable zone \citep{Yang_2013}, 
4.\ the red stellar radiation results in a weaker ice-albedo feedback and hence stabler climate \citep{Joshi_2012, Shields_2013, Shields_2014}, and 5.\ the slow main sequence evolution of M-Dwarfs means that a geological thermostat is not strictly necessary to maintain habitable conditions for billions of years \citep{Kasting_1993}.  \emph{Studying temperate terrestrial planets around red dwarfs is our best shot at understanding habitability writ large.}   

\subsection{Transit}
When a planet passes in front of its host star, it blocks a fraction of the star's light equal to the planet/star area ratio. Inferring the stellar radius from its color and surface gravity, it is possible to convert this relative size into a physical dimension. 

\begin{figure}[!htb]
\begin{center}
\vspace{-0.1cm}
      \includegraphics[width=60mm, angle=90]{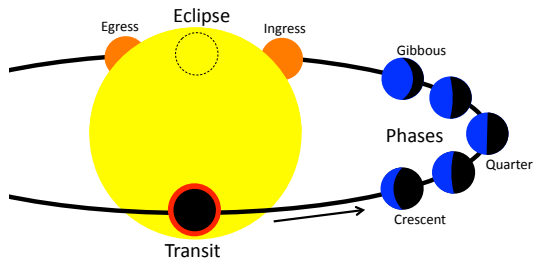}
    \caption{A planet whose orbit is nearly edge-on will \emph{transit} in front of its star.  This is more likely to occur for planets that orbit close to their star, so transiting planets are often synonymous with short-period planets. The amount of light blocked during transit tells us the planet's size, while the transmission spectrum during transit is sensitive to scattering and absorption in the planet's upper atmosphere.  During transit we see the planet's nightside, but the planet's orbital motion eventually brings the dayside into view.  The dayside reflects sunlight and is often hotter than the nightside, causing variations in brightness known as \emph{phases}. At the top of the orbit, the star will \emph{eclipse} the planet, allowing us to measure the brightness of the star without contamination from the planet; a difference measurement yields the planet's dayside brightness. Reflected light eclipse measurements are sensitive to the planetary albedo and reflected light spectroscopy constrains the nature of scattering.  Thermal eclipse measurements can be used to estimate the planet's dayside temperature, while thermal spectroscopy is sensitive to atmospheric composition and vertical temperature profile. \citep[from][]{Cowan_2014} \label{transit_cartoon}}
\vspace{-0.1cm}
\end{center}
\end{figure}

On top of the opaque planetary disk there is an annulus of partially transparent atmosphere that filters starlight (Figure~\ref{transit_cartoon}).  The spectrum of a planet in transit therefore contains an imprint of scattering and absorption that occurs in the upper atmosphere near the planet's day-night terminator. Even if the bulk composition can be taken as a given (e.g., H+He for jovian worlds), it is still difficult to nail down the abundances of trace gases, and high-altitude hazes can wash out spectral features \citep{Burrows_2014}: even if the sky looked clear to an inhabitant, it may very well be opaque to the grazing rays relevant for transit spectroscopy. Nonetheless, transmission spectroscopy is a powerful characterization tool that can only be applied to transiting planets.

\subsection{Eclipse}
It is possible to use eclipses of the planet by its star to isolate planetary light. A planet that passes directly in front of its host star usually passes directly behind it half an orbit later. The brightness of the planetary system immediately before and after occultation is compared to the brightness during eclipse, and the difference is a measure of the planet's dayside brightness. 

We may then convert the eclipse measurement into an estimate of the planet's geometric albedo or dayside brightness temperature, depending on whether the instrument is sensitive to visible or thermal radiation. Spectrally resolved eclipse measurements can constrain atmospheric scattering, composition, and vertical temperature profile \citep{Burrows_2014}. 

High-resolution emission spectra are able to resolve molecular lines (as opposed to bands), providing two novel capabilities: performing Doppler measurements on the planet itself, and probing the compositions of cloudy worlds.  For transiting planets the system inclination and planetary mass are known, but Doppler measurements might be used to infer high-altitude wind velocities \citep{Snellen_2010} and planetary rotation \citep[][]{Snellen_2014}, but these are arguably accessible via thermal phase curve and eclipse mapping measurements \citep{Rauscher_2014}, or transit morphology \citep{Carter_2010}. Clouds, on the other hand, have been most problematic in grazing transit geometry \citep{2014Natur.505...69K, 2014Natur.505...66K}. Although many emission spectra have so far been featureless, this is likely due to coarse spectral resolution, large measurement uncertainties, and roughly isothermal atmospheres \citep[][and references therein]{Hansen_2014}.

The eclipse depth is sensitive to the hemisphere-averaged properties of a planet, while the very beginning and end of an eclipse offer a means of resolving the planet's dayside:  as the planet disappears behind its star and reappears (ingress and egress, respectively), the star's edge scans across the planet. It is possible to invert these raster scans to construct a coarse two-dimensional map of the planet's dayside.  The first-order effect is the phase offset of the eclipse due to a zonally-advected hot-spot \citep{Williams_2006} and was first detected by \cite{Agol_2010}. The eclipse timing offset is largely degenerate with orbital eccentricity, but may be teased out if it is chromatic. The detailed morphology of ingress and egress provides a smaller but more robust signal about the 2D flux distribution of the planet's dayside \citep{Rauscher2007}, observed by \cite{Majeau2012} and \cite{deWit2012}.     

\subsection{Phases}
Horizontal and temporal differences in planetary temperature produce time variations in thermal emission. For example, the dayside of a slowly-rotating planet might appear warmer and hence brighter than its nightside. Most known exoplanets have curiously eccentric orbits and therefore experience significant seasons due to the changing star-planet separation.  Rocky planets are expected to form with randomly oriented spin axes, which leads to obliquity seasons. Disentangling the diurnal cycle, eccentricity seasons and obliquity seasons based on thermal phase variations is a work in progress \citep{Cowan_Voigt_Abbot_2012}.  

Tides damp obliquity, slow planetary rotation, and damp orbital eccentricity \citep[e.g.,][]{Heller_2011}, so most short period planets do not experience seasons, but probably have permanent day and night hemispheres. The day--night temperature contrast is therefore an indirect measure of atmospheric heat transport \citep{Cowan2007, Cowan_2012}. High-precision thermal phase curves can be inverted to construct coarse longitudinal temperature maps of short-period exoplanets \citep{Knutson2007, Cowan_Agol_2008}, and even contain indirect information about the planet's latitudinal flux distribution \citep{Cowan_2013}. Full-orbit observations of emission spectra enable spatially-resolved inferences of temperature structure and composition \citep{Stevenson_2014}.

With full-phase observations at high-cadence, high signal-to-noise, and high spectral resolution, it should be possible to constrain the 3-dimensional composition and temperature of an exoplanet's atmosphere.  This would allow, for the first time, realistic initialization and/or testing of a
general-circulation model with active chemistry (as opposed to fixed chemistry).  \emph{This is the
  holy grail of atmospheric characterization, and a top priority for
future exoplanet observations.}

\section{Lessons Learned}
\textbf{The principle lessons learned from the first dozen years of exoplanet atmospheric observations are: (1) short-period planets are not all alike, nor are they  a one-parameter family, (2) observations are usually systematics-limited, making out-of-occultation data critical to modeling detector behavior, and dictating that repeatability is the only reliable test of accuracy.} 

\subsection{Short-Period Planets}\label{lessons_planets}
Observations of exoplanet atmospheres have so far been weakly constraining, but they have provided a few robust surprises \citep[for more complete reviews, see][]{Burrows_2014b, Burrows_2014, Bailey_2014, Heng_2015}.   

Alkali metals have been detected in the atmospheres of some hot Jupiters \citep{Charbonneau_2002}, but are obscured by Rayleigh-scattering hazes in others \citep{Pont_2013}; a Rayleigh scattering slope has also been seen in the transmission spectrum of a warm ice giant \citep{Biddle_2014}. Hazes are present on a warm sub-Neptune \citep{2014Natur.505...69K} and a hot Neptune \citep{2014Natur.505...66K}, but absent on another \citep{Fraine_2014}.  Some hot Jupiters are slowly evaporating \citep{2004ApJ...604L..69V}, while there is questionable evidence that one planet is overflowing its Roche lobe and accreting onto its star \citep{Fossati_2010, Cowan_2012}. 

Water vapor absorption has now been securely detected in hot Jupiters with the WFC3 instrument on HST using both transit spectroscopy \citep{Deming_2013, Huitson_2013, Wakeford_2013, Mandell_2013} and eclipse spectroscopy \citep{Kreidberg_2014b}, and one hot Neptune shows evidence of water as well \citep{Fraine_2014}.  However, claims of molecular absorption in transit and eclipse measurements with other instruments remain controversial.  Early detections in NICMOS spectroscopy data \citep{Swain_2008, Swain_2009a, Swain_2009b} have been called into question \citep{Gibson_2011, Gibson_2012, Crouzet_2012}, while analyses of multi-band eclipse photometry with \emph{Spitzer} showing evidence of temperature inversions \citep{Knutson_2008}, disequilibrium chemistry \citep{Stevenson_2010, Swain_2010} and super-Solar C/O \citep{Madhusudhan_2011} have been disputed by subsequent studies \citep{Beaulieu_2011, Mandell_2011, Crossfield_2012, Zellem_2014, Diamond-Lowe_2014}.\footnote{Even when the data reduction scheme is taken at face value, disagreements may occur based on the assumptions that go into spectral retrieval. For example, while \citet{Madhusudhan2011} forces plausible chemistry,  \citet{Line2013b} and \citet{Line2014} do not. Since eclipse spectroscopy represents disk-averaged emission from regions with different temperature structures and possibly different chemistries, it is not obvious which strategy is more sound.  A pragmatic solution might be to report the best-fit ``reasonable chemistry'' solution, but with uncertainty estimates that are ``chemistry agnostic.''}   As the acquisition, reduction, and analysis of eclipse measurements have improved, the broadband emission measurements of hot Jupiters have trended toward featureless Planck spectra \citep{Hansen_2014}.

Dayside broadband emission measurements of hot Jupiters suggest that the hottest planets cannot effectively transport heat to their nightside \citep{Cowan_2011b}, and this trend has been corroborated by thermal phase measurements of two planets \citep[][]{Cowan_2012, Maxted2013}.  Thermal phase and eclipse maps are consistent with an equatorial hotspot \citep{Majeau2012, deWit2012} and suggest eastward equatorial winds at a variety of depths \citep{Knutson2007, Knutson2009, Knutson2012} with a characteristic advective time comparable to the radiative timescale \citep{Agol_2010}. Thermal phase measurements of hot Jupiters on eccentric orbits suggest radiative times less than a day and equatorial super-rotation \citep{Lewis2013, Wong_2014}.  

The measured geometric albedos of transiting planets are generally small \citep{Rowe_2008, Kipping_2011} and blue \citep{Evans_2013} but with notable exceptions \citep{Demory_2011}.  Moreover, the one hot Jupiter with high albedo appears to have spatially inhomogeneous clouds \citep{Demory_2013}.  

\subsection{\emph{Spitzer}}
By operating in the mid-infrared without
the confounding telluric IR background, the Spitzer Space Telescope
overcame a key barrier and measured the first exoplanetary photons in
the Fall of 2004.  Teams led by \citet{Charbonneau2005} and
\citet{Deming2005} observed secondary eclipses of TrES-1 and HD
209458b, respectively, submitting their independent reports
simultaneously and participating in a joint NASA press release in
April 2005.  \emph{Spitzer} remained the observational tool of choice for most of the subsequent decade, and it is useful to summarize what we have learned during this time.  

In its cryogenic mission (2003 -- 2009), \emph{Spitzer} had six
photometric channels useful for exoplanet characterization, centered
at 3.6, 4.5, 5.8, 8, 16, and 24 \micron.  Channels ranged from 1--3~\micron~wide.  \emph{Spitzer}'s InfraRed Spectrograph had a
low-resolution, 5.3--14~\micron~mode useful for exoplanets.
Unfortunately, the rate of transiting-planet detection was low until
around 2010, when the cryogen was already gone.  In Warm \emph{Spitzer} only
the 3.6 and 4.5 \micron~photometric channels remain operational; spectroscopy
is unavailable.  Thus, only HD~209458b and HD~189733b were ever
observed spectroscopically; only they and GJ 436b were observed at 24
\micron; and only GJ~436b, TrES-1, TrES-4, HD~189733b, and
HD~149026b were observed at 16 \micron.  However, numerous planets
have measurements in the four shortest bands, and even more have published 3.6 and 4.5 \micron~eclipse depths \citep{Hansen_2014}. 

With an aperture of just 85~cm, \emph{Spitzer} was designed for 10\% absolute
and 1\% relative photometry, an order of magnitude poorer precision
than required for exoplanet studies.  The race to achieve results thus
became a contest of systematic correction methods.  Bayesian sampling (e.g., Markov Chain Monte Carlo)
replaced model fitting by simple $\chi\sp{2}$ minimization, Bayesian
methods for comparing models with different numbers of free parameters
became the norm, and a variety of numerical approaches modeled
\emph{Spitzer}'s systematics.  Although known for a decade or more, improved
image centering and photometric extraction techniques also finally
entered common use.  Ultimately, contrast uncertainties better than
0.01\% for single eclipses became possible.  While an impressive
improvement over \emph{Spitzer}'s design, the 0.01\% uncertainty still left
the broadband measurements at S/N of only $\sim10$ for most observed
planets. The 1 -- 3 \micron~width and small number of photometric channels
further confounded spectroscopic retrieval.

Most \emph{Spitzer} exoplanet observations consist of unrepeated eclipse observations obtained with different observing schemes and analyzed by disparate groups using a variety of evolving reduction pipelines.  The poor signal-to-noise is mostly a testament to the relative faintness of currently known transiting planets; NASA's TESS mission will address this problem. The observations, however, would have greatly benefited from instruments designed to stare at bright point sources, and the same observations would have had much more scientific value had they been obtained in a uniform way and reduced with a uniform, open-source pipeline. \emph{The \emph{Spitzer} experience ultimately justifies the
expense of a purpose-built mission to perform a survey of exoplanet atmospheres.}   

\emph{Spitzer}'s after-launch data analysis efforts attempted
to characterize the varying sensitivity across the faces of individual
pixels and to fit its temporal response curves.  However, pre-launch
laboratory calibration measurements at better than the 0.01\% level
could have done a better job, as could calibrating long stares at
point sources.  For example, the prominent, time-dependent sensitivity
``ramp'' at 5.8, 8, and 16 \micron~was unknown prior to launch
because lab calibrations on bright sources lasted only a short time,
rather than the hours-long timescale of the ramp.  Learning from this
experience, instruments that reduce intrapixel sensitivity variations,
that have long-term stability, and that point consistently are now
being designed and proposed for dedicated missions.  Even JWST, which
was designed before some key lessons were learned from \emph{Spitzer}, is
being calibrated with enhanced emphasis on observing modes suitable
for exoplanet studies.

\section{Exoplanet Observatories Through 2025}
\textbf{Many ground- and spaced-based observatories can be used to study the atmospheres of hot Jupiters and warm sub-Neptunes in the next decade.  For terrestrial planets, especially the temperate variety, the choice of instruments is much more limited: only the Extremely Large Telescopes and the James Webb Space Telescope will be capable of atmospheric observations.}  

\subsection{4m telescopes} 
4-meter class telescopes are an underused resource in the atmospheric characterization of exoplanets.  Ground-based, optical multi-object spectrographs typically capture light from 0.4--1.0 microns and are sensitive to the presence of H$_2$O, alkali metals such as Na and K, and metal hydrides in cloud-free atmospheres.  Additionally, in atmospheres with clouds or hazes, we can use spectral information to determine the size distribution and altitude of the cloud/haze particles.
   
In order to achieve the equivalent precision and be competitive with larger telescopes, these smaller telescopes must acquire multiple transits of a single target.  For comparison, four transits of a single target with a 4-meter telescope, such as Mayall or Blanco, have the equivalent photon-limited precision of two transits with an 8-meter telescope, such as Gemini.  However, acquiring numerous transit observations with the smaller telescope will mitigate residual atmospheric effects that limit our precision with ground-based observations, plus these observations will likely have a higher duty cycle, thereby outperforming the larger telescope.  With a sufficient number of transits, 4-meter class telescopes can contribute exciting, cutting-edge research like with larger telescopes, but at a fraction of the cost.

High precision photometry in the near-infrared of exoplanet occultations provides a direct measurement of thermal emission from hot planets, with single-eclipse precisions of $2\times10^{-4}$ now all but routine \citep{Croll_2014}.  \emph{Wide-field NIR photometry with 4m class telescopes is a viable means of achieving high precision photometry for gas and ice giants orbiting relatively bright stars}.  The wide field of view allows for a significant number of reference stars for differential photometry, which is critical for minimizing effects of Earth's atmosphere (typically the limiting factor for high precision photometry in the near-infrared from the ground).  

Despite the advantages of these types of facilities, only a handful exist: WIRCam on the 3.6m CFHT, WFCAM on the 3.8m UKIRT, NEWFIRM on the KPNO 4m Mayall Telescope, and the Spartan IR Camera on the 4.1m SOAR.  Of these, only WIRCam on CFHT and WFCAM on UKIRT have been demonstrated for exoplanet studies (e.g., Croll et al. 2011, Col\'on \& Gaidos 2013).

\subsection{10m telescopes} 
As demonstrated by \cite{Colon_2009}, large ground-based telescopes are capable of contributing significantly to photometric follow-up efforts for both small, long-period planets discovered by missions like Kepler and for larger, Jupiter-size planets.  Indeed, the first ground-based detection of sodium in an exoplanet atmosphere came from the Hobby-Eberly Telescope \citep{Redfield_2008}.  Current 10m-class facilities include the Keck telescopes, the Gran Telescopio Canarias (GTC), Hobby-Eberly Telescope, Southern African Large Telescope, Subaru, Very Large Telescope, Large Binocular Telescope, and Gemini.  These facilities offer a number of instruments suitable for exoplanet atmospheric characterization.  For example, the primary instrument on the GTC suitable for exoplanet observations is the Optical System for Imaging and low Resolution Integrated Spectroscopy \citep[OSIRIS]{Cepa_2000,Cepa_2003}, which offers a moderately sized field of view of $7.8 \times 7.8$~arcmin.  OSIRIS has the capability for standard long-slit spectroscopy, and it also offers a unique tunable filter imaging mode, allowing the user to specify custom optical bandpasses with FWHM = 1.2--2.0 nm.  In this mode, observations can be conducted in multiple tunings nearly-simultaneously, thus allowing for narrow-band spectrophotometry.  \cite{bean:2010} used multi-object spectroscopy on the VLT to obtain transit spectra of GJ~1214b, with an epoch-to-epoch white-light accuracy of $\sim2\times10^{-4}$. Other up-and-coming facilities include GMOS on Gemini and MOSFIRE on Keck.  \emph{Thanks to their large aperture 10~m telescopes are capable of the high precision spectroscopy or spectrophotometry required to measure the small signals from exoplanet atmospheres, e.g., alkali metals in hot Jupiter atmospheres \citep{Sing_2011} or methane in warm sub-Neptunes \citep{Wilson_2014}}.

\subsection{Extremely Large Telescopes} 
By 2025, we expect 1--3 of the extremely large, ground-based
telescopes (ELTs) currently being planned (GMT, TMT, E-ELT) to be
operational. These facilities may provide several possible benefits:
their larger apertures will enable studies of larger numbers of
transiting planets around fainter stars, with techniques used today on
8--10~m telescopes; if systematic effects can be controlled, the
larger apertures may also allow planets in brighter systems to be
studied at higher precision; finally, by relieving some pressure on
existing 8--10~m telescopes it may be possible to use these older
facilities for large-scale survey science precluded today due to the
limited availability of observing time.

The ELTs will be optimized to observe faint targets, often using
adaptive optics -- implying relatively narrow fields of view. Thus
fewer nearby systems will be accessible to the differential photometric and
spectroscopic techniques so popular today, which use multiple
comparison stars to remove telluric and instrumental
noise \citep{bean:2010,gibson:2013,crossfield:2013b}.  Today, these
efforts perform within factors of 2--3 of the spectroscopic photon
noise limit, but there is no consensus as to whether the extra noise
comes from detectors, instruments, telescopes, or the Earth's
atmosphere. \emph{For ELT multi-object transit spectroscopy to succeed,
these noise sources must be identified and mitigation strategies
incorporated into the new instruments currently being designed.}

High-dispersion spectroscopy may be the niche in which ELTs can best
make significant progress. Observations at high dispersion
($\lambda/\Delta \lambda \gtrsim 20,000$) can measure unique molecular
signatures, thermal profiles, global atmospheric circulation, and
orbital motion and has already been used to characterize the
atmospheres of several transiting planets
\citep{snellen:2010,crossfield:2011,rodler:2012,birkby:2013}. Such observations
also come within 20\% of the photon noise limit \citep{brogi:2014},
which indicates the technique is less susceptible to the systematic
effects that limit multi-object observations. Thus, ELT
high-dispersion spectroscopy may be especially well-suited to high-precision
atmospheric studies of transiting planets.  

\emph{High dispersion spectroscopy with the ELTs will be also key to detecting atmospheric
signatures of transiting Earth-like planets hosted by nearby stars, to be discovered by TESS and
PLATO} \citep[see][]{snellen:2013,rodler:2014}.

\subsection{SOFIA}
SOFIA combines a number of advantages for extremely precise time-domain optical and near-infrared spectrophotometric observations using its HIPO \citep{2004SPIE.5492..592D} and FLITECAM \citep{2006SPIE.6269E.168M} instruments, in particular when used in simultaneous FLIPO mode \citep{2010PASP..122.1020A,2012SPIE.8446E..18NS}.
SOFIA is able to avoid most of the perturbing variations of atmospheric trace gases that are the main source of systematic noise for ground-based observations at shorter NIR wavelengths, as these telluric molecules are also the species of interest in the exoplanetary atmospheres. The SOFIA telescope, operating at much lower temperatures (240~K) than ground-based telescopes, reduces thermal background contributions that are the dominant noise source for transit observations at wavelengths longer than 3 microns.  SOFIA can observe time-critical events, such as planetary occultations, under optimized conditions, as demonstrated by an observation of a Pluto occultation in June 2011 \citep{2013AJ....146...83P}. SOFIA's initial forays into transit observations have yielded precisions within a factor of 2 of the Poisson limit (Angerhausen et al.\ submitted).  \emph{In contrast to space telescopes it is possible to update SOFIA with state-of-the-art instrumentation}, e.g.\ with a dedicated 2nd generation exoplanet instrument  \citep[such as NIMBUS,][]{2012SPIE.8446E..7BM}.

\subsection{Hubble Space Telescope}
Planetary emission peaks in the near-to-mid-infrared, while stellar
emission falls dramatically longward of its peak in the visible to
near-infrared.  In addition,
prominent molecular rotational-vibrational bands for numerous abundant
molecules, such as water, methane, carbon monoxide, and carbon
dioxide, occur in the near- and mid-infrared.  The IR has thus become
the most productive spectral region for exoplanet characterization,
despite the disadvantages of poorer technology, orders-of-magnitude
worse thermal background, and fewer photons compared to the visible
range.

HST/WFC3 can be used for NIR transit spectroscopy of any planet with large scale-height atmosphere, and emission spectroscopy for the hottest planets. It has proven capable of 30~ppm spectrophotometry \citep[in fifteen 7-pixel spectral bins;][]{2014Natur.505...69K}, sufficient to detect molecular features (\S\ref{lessons_planets}).  While the limited wavelength range prohibits detailed atmospheric retrieval using HST measurements alone, it is sufficient to determine which exoplanets are hazy. It is also possible to stitch together multiple HST orbits to obtain continuous phase measurements of an exoplanet, hence constraining its global albedo and heat transport \citep{Stevenson_2014}.

HST/COS can be used to perform reflection spectroscopy of the shortest-period giant planets \citep{Evans_2013} as well as to characterize the UV environment of temperate terrestrial planets, and hence photochemistry of water, carbon dioxide, methane, oxygen, and ozone \citep{Linsky_2013, France_2014}. \emph{As the only UV instrument available in the forseeable future, a high priority for HST must be to characterize the stellar UV spectrum of temperate terrestrial transiting planets.} This will be critical in assessing the habitability of these worlds \citep{Wordsworth_2013} and the eventual detection of biosignatures \citep{Tian_2014}. Since HST and its UW spectroscopic instruments, COS and STIS, have a limited lifetime ($\sim$2019), \emph{the need to obtain UV spectra of exoplanet host stars and transmission spectra of planets may justify a future UV mission}.

\subsection{K2, TESS, CHEOPS, PLATO}
\emph{These space-based transit search telescopes can measure white-light geometric albedos and scattering phase functions for hot planets, but with unknown contamination from the planet's thermal emission} \citep[][]{Cowan_2011b, Esteves_2014}.  Even a simple red vs.\ blue bandpass, as was available on CoRoT, would greatly alleviate this problem.  Alternatively, white-light optical measurements can be combined with thermal measurements obtained with other observatories in order to obtain geometric albedo estimates \citep[e.g., \emph{Spitzer};][]{Demory_2013}. 

The geometric albedo constrains atmospheric scattering, and can in principle be converted into a Bond albedo to inform climate calculations, but only by making assumptions about the planet's reflectance spectrum, spatial homogeneity, and scattering phase function.

\subsection{Spitzer Space Telescope}
The \emph{Warm Spitzer} mission has two functioning IRAC channels at 3.6 and 4.5 \micron, with a demonstrated precision of $<10^{-4}$ with intensive campaigns. These are suitable for broadband transit and eclipse measurements, but on its own, \emph{Spitzer} only provides radius and temperature measurements for transiting planets. On the other hand, \emph{Spitzer's ability to continuously stare at a planetary system is unique}.  Moreover, the mid-IR wavelengths observable with \emph{Spitzer} help anchor the continuum level of transit and eclipse spectra \citep{Fraine_2013}.
 
\subsection{James Webb Space Telescope}
The James Webb Space Telescope (JWST) has four instruments, all of which can be used to study transiting exoplanets: Near-Infrared Camera (NIRCam), Near-Infrared Spectrograph (NIRSpec), Near Infrared Imager and Slitless Spectrograph (NIRISS), and Mid-Infrared Instrument (MIRI). For detailed discussion of the instruments and modes best suited to exoplanet studies, see \cite{Beichman_2014}. One of the key takeaways from that report is that although JWST has sensitivity from 0.6--28~\micron, at least four passes with different instruments/modes will be necessary to obtain a 1--12~\micron~spectrum for nearly all systems (faint targets can be observed in 2 passes, provided they don't saturate NIRSpec near 1--2~$\mu$m). \emph{JWST will be the most powerful space observatory for transiting planets in the coming decade, and the only one able to characterize the climate of temperate, terrestrial planets}. 

\section{Benchmark Planets}\label{benchmarks}
\textbf{JWST and ELTs will easily characterize the atmospheres of nearby hot Jupiters, hot super-Earths, and warm sub-Neptunes. However, atmospheric measurements of temperate rocky planets transiting nearby M-Dwarfs will be much harder: (1) detection of reflected light or detection of the Planck peak of planetary emission is impossible in a single occultation, (2) transit spectral features in atmospheres free of clouds/hazes are barely detectable in a single occultation with JWST; such measurements could be feasible with ELTs if they approach photon-counting precision in the optical--NIR, and (3) planetary emission on the Rayleigh-Jeans tail should be barely detectable with JWST/MIRI in a single occultation.  Since JWST cannot achieve a robust atmospheric measurement of an Earth-analog in a single occultation in the photon-counting limit, characterizing such planets will require campaigns lasting weeks to months.}

We consider four fiducial transiting planetary systems (Table~\ref{parameters}): a hot Jupiter, a hot super-Earth, a warm sub-Neptune, and a temperate super-Earth.  The two ``hot'' planets orbit K-dwarfs, while the cooler planets orbit M-dwarfs.

\begin{deluxetable}{lcccrrrrr}
\tabletypesize{\scriptsize}
\tablecaption{Fiducial Transiting Planet Parameters \label{parameters}}
\tablewidth{0pt}
\tablehead{
\colhead{Planet Type} &\colhead{Stellar}& \colhead{Stellar} & \colhead{Planetary}& \colhead{\emph{Transit}}& \colhead{$a/R*$}&\colhead{\emph{Dayside}}& \colhead{Mean Mol.}&\colhead{Planet}\\
 &\colhead{Temp.}& \colhead{Radius} & \colhead{Radius}& \colhead{\emph{Depth}$^{a}$}&&\colhead{\emph{Temp.}$^a$}& \colhead{Mass}&\colhead{Gravity}
}
\startdata
Hot Jupiter &5000~K& $0.75R_\odot$ & 1.10~$R_J$& $2.2\times10^{-2}$&5& 1787~K& 2~$\mu$&20~m/s$^2$\\
Hot Super-Earth &5000~K& $0.75R_\odot$& 1.50~$R_\oplus$ &$2.9\times10^{-4}$&3&2308~K& 140~$\mu$&12~m/s$^2$\\
Warm Sub-Neptune &3000~K& $0.20R_\odot$& 0.24~$R_J$& $1.4\times10^{-2}$&15&619~K&2~$\mu$ &9~m/s$^2$\\
Temperate Super-Earth &3000~K& $0.20R_\odot$& 1.50~$R_\oplus$ &$4.1\times10^{-3}$&90&253~K& 28~$\mu$&12~m/s$^2$ 
\enddata
\tablecomments{$^a$Derived quantity.}
\end{deluxetable}

\subsection{Signal}
The reflected light contrast is simply $[F_p/F_*]_{\rm ref} = A_g(R_p/a)^2$, where $A_g$ is the geometric albedo (assumed to be 0.3 for all four planets), $R_p$ is the planetary radius, and $a$ is the semi-major axis.

The amplitude of transit spectral features is approximated as $2R_p N_H H/R_*^2$, where the stellar radius is $R_*$, the atmospheric scale height of the planet  is $H = k_B T/(\mu g)$, atmospheric temperature is $T$, atmospheric mean molecular mass is $\mu$, and surface gravity is $g$.  The number of scale heights probed, $N_H$, is approximately 4 \citep{Griffith_2014}.  We set the atmospheric temperature to the dayside effective temperature, described below.

The dayside emitting temperature depends on the incident stellar flux, the fraction that is reflected away, and the fraction that is transported to the planet's nightside. Adopting the parametrization of \cite{Cowan_2011b}, we have $T_d = T_* \sqrt{R_*/a} (1-A_B)^{1/4} (2/3 - 5\varepsilon/12)^{1/4}$, where $T_*$ is the stellar effective temperature, and we assume a Bond albedo of $A_B=0.3$ and heat recirculation efficiency of $\varepsilon=0.2$ for all planets.

If we treat the planet and its host star as blackbodies, then the thermal contrast ratio between them is proportional to the ratio of Planck functions: $[F_p/F_*]_{\rm therm} = (R_p/R_*)^2 B(\lambda, T_d)/B(\lambda, T_*)$.  The peak of the planetary emission is given by Wien's Law, $\lambda_{\rm peak} = 2898/T_d$, while in the Rayleigh-Jeans limit ($\lambda \to \infty$) the thermal contrast is simply $(R_p/R_*)^2 (T_d/T_*)$.

\begin{deluxetable}{lrrrr}
\tabletypesize{\scriptsize}
\tablecaption{Atmospheric Signals \label{values}}
\tablewidth{0pt}
\tablehead{
\colhead{Planet Type} & \colhead{Reflected} &\colhead{Transit}  & \colhead{Planet Peak} & \colhead{Rayleigh-Jeans}\\
& \colhead{Contrast}  &\colhead{Feature} & \colhead{Contrast}& \colhead{Contrast}
}
\startdata
Hot Jupiter & \boldmath{$2.6\times10^{-4}$} & \boldmath{$8.3\times10^{-4}$} &\boldmath{$7.4\times10^{-4}$}&\boldmath{$7.7\times10^{-3}$}\\
Hot Super-Earth & \boldmath{$9.8\times10^{-6}$} &\boldmath{$3.0\times10^{-6}$}& \boldmath{$1.8\times10^{-5}$}& \boldmath{$1.4\times10^{-4}$}\\
Warm Sub-Neptune &  $1.9\times10^{-5}$ &\boldmath{$2.0\times10^{-3}$}& $1.8\times10^{-4}$& \boldmath{$3.0\times10^{-3}$}\\
Temperate Super-Earth & {\color{red}$1.5\times10^{-7}$} &$2.3\times10^{-5}$& {\color{red}$1.5\times10^{-5}$}& $3.5\times10^{-4}$ 
\enddata
\tablecomments{Fonts denote the difficulty of measuring these signals with JWST in the white-light \emph{photon-counting limit} (cf.\ Table~\ref{precisions}): \boldmath{bold} indicates signal measured at $>10\sigma$ in a one hour integration, while {\color{red}red} denotes $<1\sigma$.}
\end{deluxetable}

The amplitudes of atmospheric signatures for the four planets are listed in Table~\ref{values}. If one simply considers the amplitudes of the atmospheric signals, the easiest measurement is the thermal secondary eclipse, which constrains a planet's emitting temperature, followed by emission spectroscopy to constrain vertical temperature structure.\footnote{A typical temperature contrast of $0.5^{1/4}\approx0.8$ between the infrared photosphere and the stratosphere produces molecular emission features of order unity.} For hot Jupiters and hot Earths, thermal phase variations---constraining horizontal temperature structure---are similar in amplitude to eclipses, but require much longer observations. With the exception of the warm sub-Neptune, transit spectroscopy---yielding atmospheric scale height and opacity---is a smaller signal, and in practice requires higher spectral resolution.   The smallest signal is typically reflected light, which informs atmospheric scattering and clouds. 

It is worth discussing what habitability constraints such measurements would provide for the handful of transiting temperate terrestrials TESS is expected to discover within the decade. The detection of planetary emission places a lower limit on the surface temperature of the planet (modulo the greenhouse effect), while broadband thermal phase measurements can be used to infer a planetary atmosphere \citep{Seager_2009}, constrain its mean-molecular weight and opacity \citep{Menou_2012, Kataria_2014} and surface pressure \citep{Koll_2015}, and might indicate the presence of water clouds \citep{Yang_2013}. Measuring a transit spectral feature, on the other hand, will establish that the planet has an atmosphere and signal the presence (but not abundance) of a molecule at low pressures \citep{Burrows_2014}. With a full transit spectrum from the optical through the infrared it would be possible to uniquely determine the composition of such a planet's upper atmosphere \citep{Benneke_2012}.

\subsection{Noise}
While the atmospheric signal depends on a combination of planetary and stellar parameters, the overall flux---and hence noise---is dominated by the star. We compute the number of photons collected by an instrument, accounting for limited bandpass and imperfect system throughput:
\begin{equation}
N_{\rm phot} = \frac{\pi \tau \Delta t}{hc}\left(\frac{R_* D}{2d}\right)^2 \int_{\lambda_1}^{\lambda_2} B(\lambda, T_*) \lambda d\lambda ,
\end{equation}
where $\lambda_1$--$\lambda_2$ is the bandpass, $d$ is the distance to the star (fixed to 20~pc), $D$ is the telescope diameter, $\tau$ is the system throughput (photon conversion efficiency = electrons out per photon in), and $\Delta t$ is the integration time (set to 1~hour). The telescope and instrument parameters used are listed in the top half of Table~\ref{precisions}.  In the poisson limit for large numbers of photons, the precision is simply $1/\sqrt{N_{\rm phot}}$, as listed in the bottom half of Table~\ref{precisions}.

It is important to note that the numbers listed in Table~\ref{precisions} are \emph{white light} precisions and therefore only directly apply to measurements of reflected light, thermal emission, or broadband spectroscopy.  The precision for spectroscopy is less impressive.  For example, if one wanted to obtain a 1.0--4.0~$\mu$m spectrum with 0.1$\mu$m resolution using JWST/NIRSpec, then the expected precision would be $\sqrt{30}\approx 5.5$ times worse than the values shown in the table.  

The amplitude of spectral features for a temperate terrestrial planet transiting in front of a nearby M-Dwarf is therefore comparable to the single-epoch photon-counting precision.  In the absence of a systematic noise floor, 100 transits of such a planet could yield 10$\sigma$ detections of greenhouse gases.  Spending a total of one month of JWST time to characterize the atmosphere of a potentially habitable world is compelling, but the observations would have to be spread out over nearly a decade for a planet in a month-long orbit (this scheduling problem is alleviated for planets in the habitable zones of later M-Dwarfs, which have shorter orbital periods).

Table~\ref{precisions} suggests that detecting thermal emission from temperate terrestrials is a better option, but one should be cautious here, too: at longer wavelengths, the dominant source of noise may not be photon counting, but the warm detector background. Moreover, staring continuously at an M-Dwarf to monitor the phase variations of a temperate planet poses technical challenges that have not yet been resolved.  \emph{Any atmospheric measurement of a temperate terrestrial planet will be difficult for JWST, and any robust atmospheric signature will require at least a month of observing time.}

\subsection{Comparison to Empirical Precisions and Other S/N Estimates}\label{comparison}
Since these precision estimates are idealized, it is worth considering a few concrete examples to make sure they are approximately correct.  

\cite{2014Natur.505...69K} report 30~ppm ($3\times10^{-5}$) precision for transits of GJ1214b, at the photon-counting limit.  This planet is closer than our nominal warm sub-Neptune (13 instead of 20~pc), the authors observed 12 transits lasting 0.9 hours each, and divide the HST/WFC3 G141 band into 22 spectral bins. The eclipse depth is differential measurement so the uncertainty in the out-of eclipse baseline contributes to the overall uncertainty, a $\sqrt{2}$ penalty.  Combining these adjustments, we would naively expect a photon-limited precision of $(1.6\times10^{-5})(13/20)\sqrt{22/(12\times0.9)}\sqrt{2} = 2\times10^{-5}$.

\cite{Knutson2012} report precisions of $5\times10^{-5}$ for a single eclipse of HD~189733b measured with \emph{Spitzer}/IRAC ch1, a factor of a few worse than the Poisson limit. Since this planet is nearly identical to our notional hot Jupiter, we can compare to our predicted photon limit for the same instrument (Table~\ref{precisions}). The eclipse duration is $1.8$ hours, and we again account for the $\sqrt{2}$ penalty for a differential measurement with equal number of photons in and out of transit.  These two effects roughly offset, so Spitzer/IRAC should be able to measure a single eclipse of this planet with a precision of approximately $2\times10^{-5}$, in the photon-counting limit. 

\cite{Crossfield_2012} report a precision of $3\times 10^{-4}$ for three eclipses of HD~209458b observed with \emph{Spitzer}/MIPS at 24~$\mu$m. We begin with the value of $8.4\times 10^{-5}$ from Table~\ref{precisions}. Ignoring the somewhat hotter star (6000 rather than 5000~K) but accounting for the more distant system (50~pc), $3\times3$~hour eclipses, and the $\sqrt{2}$ penalty for a differential measurement, we obtain a photon-counting precision of $9.9\times10^{-4}$, a factor of three better than the empirical accuracy obtained by \cite{Crossfield_2012}. 

In short, the values listed in Table~\ref{precisions} accurately reflect the best-case scenario of photon counting precision.

We can also compare our JWST precision estimates to those made by other groups.  Detailed simulations of JWST/NIRSpec indicate that it will only be possible to detect water absorption in transit spectra for planets that are hydrogen-rich and/or hotter than 400~K (N.~Batalha, private communication).  In other words, \emph{the only way that JWST can easily perform transit spectroscopy of temperate terrestrial planets is to stretch the definition of the latter}. Quantitatively, they find that observing 40 transits might enable the $15\sigma$ detection of water vapor absorption for a bona fide temperate terrestrial planet.  If we begin with our JWST/NIRCam F115W estimate, and account for the approximately 120~hrs of in-eclipse time, $\sim$20 resolution elements, and 3~pc system distance, we obtain a precision of $1.3\times10^{-5}\times\sqrt{20/120}\times(3/20) = 8\times10^{-7}$, or a 14$\sigma$ detection, in excellent agreement with the estimate of Batalha et al.

\cite{Deming_2009} estimate that JWST/MIRI can obtain a 3.2$\sigma$ detection of a HZ super-Earth's Planck peak emission at 13~pc in 4$\times$90 min eclipses. If we take our JWST/MIRI F1000W estimate and account for the 6~hrs of integration, and the closer planet, we obtain an estimate of $2.6\times10^{-5}\times(13/20)\times(1/\sqrt{6}) = 7\times 10^{-6}$.  Since the signal at this wavelength is $1.5\times10^{-5}$, we would have predicted a 2$\sigma$ detection.  This is excellent agreement given the somewhat different assumptions about stellar and planetary properties. 

\cite{Yang_2013} predict a 49$\sigma$ detection of planetary emission in a 1-day integration of a $2R_\oplus$ super-Earth orbiting a 3000~K M-dwarf 20~pc away, but assuming one integrates 10--28 micron, which would take multiple passes with JWST/MIRI.  Starting with our JWST/MIRI F2550W precision estimate and accounting for the longer integration time yields $9.7\times10^{-5}/\sqrt{24} = 2\times10^{-5}$.  This would be an 18$\sigma$ detection of planetary emission according to Table~\ref{values}, but the planetary flux at 25~$\mu$m is a few times smaller than in the Rayleigh-Jeans limit. \emph{We therefore find that measuring thermal emission with JWST/MIRI is the easiest atmospheric measurement one can make for temperate terrestrial planets, but it is harder than stated in \cite{Yang_2013}.} 

One may fear that detector systematics will stymie our ability to detect the subtle signatures of temperate terrestrial planets.  This is a legitimate worry, since many of the instruments on JWST have heritage in instruments on HST and \emph{Spitzer}.  It is therefore encouraging that the current broadband precision record holders with NASA's great observatories are comparable to the amplitude of atmospheric signatures on transiting temperate terrestrials: $5\times10^{-6}$ in the NIR \citep{Stevenson_2014}, $5\times10^{-5}$ at 3.6~$\mu$m \citep{Knutson2012}, and $3\times10^{-4}$ at 24~$\mu$m \citep{Crossfield_2012}. \emph{Robustly detecting the atmospheres of temperate terrestrial exoplanets requires less than an order-of-magnitude improvement in our detector models} \cite[this is in line with the claim that the transit spectral signature of a temperate terrestrial planet could be measured with an intensive HST campaign;][]{2014Natur.505...69K}.    

\section{How to Spend JWST Time}
\textbf{During a 5~year lifetime, JWST could determine the vertical temperature structure of 150 transiting giant planets, map the atmospheres of 25 giants, \emph{or} constrain the habitability of 3 temperate terrestrial planets. A balanced portfolio of these three categories would maximize science returns.}

As a ballpark estimate based on experience with the Hubble and Spitzer Space Telescopes, we anticipate that 25\% of JWST time will be devoted to transiting planets. Multiplying this by the estimated 70\% duty cycle of the observatory and a 5 year mission, yields 320 days (should the mission last 10~years, there would be approximately 600~days to devote to transiting exoplanets).  How should we spend it? We summarize a few endmember observing portfolios in Table~\ref{portfolios} and discuss the topic in more detail below \citep[for a qualitative discussion see][]{Heng_2015}.  We focus only on the most time-consuming observations; there will undoubtedly be rare opportunities that will be pursued (e.g., temperate transiting gas giants).

\begin{deluxetable}{lccc}
\tabletypesize{\scriptsize}
\tablecaption{JWST Transiting Planet Observing Portfolios \label{portfolios}}
\tablewidth{0pt}
\tablehead{
\colhead{Portfolio Name} & \colhead{Number of Targets} & \colhead{Duration per Target} & \colhead{Total Time}}
\startdata
Atmospheric Structure & 150 & 2 days & 300 days\\
Atmospheric Mapping & 25 & 12 days & 300 days\\
Temperate Terrestrials & 3 & 100 days & 300 days
\enddata
\tablecomments{Endmember portfolios for transiting planet science with JWST.  Linear combinations of these are possible, and probably more scientifically productive: e.g., $70\times2=140$~days of structure, $5\times12=60$~days of mapping, and 1 terrestrial (100~days) also adds up to 300~days.}
\end{deluxetable}

\subsection{Hot Jupiters and Warm Sub-Neptunes}
At six hours per transit  (2 out + 2 in + 2 out) $\times$ 4 instruments (the JWST Science Working Group should codify these modes), it takes roughly
24 hrs to get a 1--10~$\mu$m transit spectrum.  Eclipse
spectroscopy will take just as long \citep[a single pass is also sufficient for eclipse mapping of the brightest targets;][]{Rauscher2007}.  \emph{It would therefore take 2 days to get a baseline atmospheric characterization of a transiting exoplanet with JWST}. If the planet's atmosphere is well mixed, one can roughly think of the transit spectra as constraining the atmospheric opacity, which allows the dayside emission spectroscopy to more robustly constrain the temperature structure. If one performed a systematic survey using JWST, it would be possible to obtain 1--10~$\mu$m transmission and eclipse spectra for 150 short-period giant planets over the course of a 5~yr mission.  

Full phase coverage constrains the planet's energy budget, heat transport, radiative response, and enables spatially-resolved chemical mapping \citep{Stevenson_2014}.  If one considers planets with a typical orbital period of 3 days, it takes 12 days to obtain a full phase-resolved spectrum of a single planet. A systematic survey using all of JWST's transiting planet time could perform such detailed observations for 25 short-period planets.

It should be noted that even a partial planetary spectrum with a single JWST instrument would be much more valuable than a pair of broadband \emph{Spitzer} measurements (currently the typical state of knowledge for hot Jupiters): the high signal-to-noise ratio and spectral resolution should enable meaningful inferences about atmospheric structure and composition. 

\subsection{Hot Earths}
TESS is expected to discover 4--6 hot Earths orbiting V=9--11 stars (Sanchis-Ojeda, priv.\ comm.).  \cite{Samuel_2014} estimates that NIRSpec observations of CoRoT-7b (V=12) would take 70~hrs, while a V=8 target would only take 3~hrs; our S/N estimates in Table~\ref{values} are in broad agreement. Given the short orbital periods, it is both expedient and scientifically important to obtain full phase curves.  Devoting 4 days each with 3 instruments (too faint for MIRI) for all 5 bright targets adds up to 60 days for exo-geology. 

\subsection{Temperate Terrestrial Planets}
Given the planetary demographics of \emph{Kepler}, TESS should deliver 4 temperate terrestrial planets ($0.5R_\oplus<R<1.5R_\oplus$ and $0.5S_\oplus<S<1.5_\oplus$, where $R$ and $S$ are planetary radius and insolation, respectively) that transit bright M-dwarfs in the JWST continuous viewing zone (P.~Sullivan, private communication).\footnote{\cite{Gaidos_2014}, on the other hand, estimate that TESS only has a 1.3\% chance of detecting a HZ planet around one of the $\sim3000$ \emph{brightest} late-K and early-M stars (CONCH-SHELL stars).}  TESS host stars are early-M so the HZ corresponds to month-long orbits. HZ worlds discovered by the MEarth survey would be better because the host stars might be cooler, the transit depths more favorable, $F_p/F_*$ larger, and the HZ orbits shorter.  But at 1.9--2.6$R_\oplus$ in radius, these planets are likely temperate sub-Neptunes (Z.~Berta-Thompson, private communication). 

Given the numbers in Tables~\ref{values} and \ref{precisions}, it would take a few months to obtain high-precision JWST/NIRSpec transit spectra and JWST/MIRI phase curves for a single nearby temperate terrestrial planet. Using all 300~days of anticipated JWST transiting planet time, we could obtain transit spectra  and thermal phase curves for 3 terrestrial HZ worlds orbiting nearby M-Dwarfs.

\subsection{Target-Limited vs.\ Time-Limited Regimes}
\begin{figure}[!htb]
\begin{center}
\vspace{-0.1cm}
      \includegraphics[width=70mm]{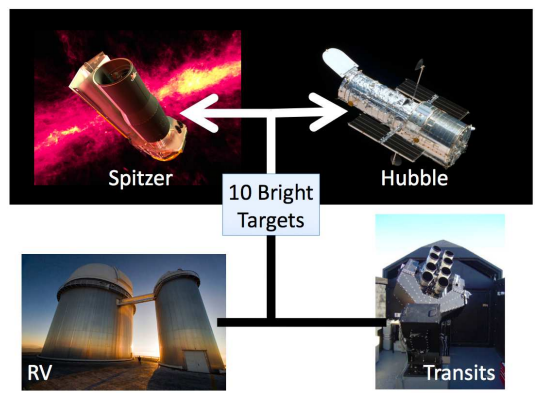}  
    \caption{\emph{Target-Limited Regime:} Past and current ground-based radial velocity and transit surveys have combined to discover and characterize hundreds of transiting exoplanets, only a dozen of which are bright enough to warrant detailed follow-up with the Spitzer and Hubble Space Telescopes. In effect, any bright transiting planet is the subject of detailed atmospheric follow-up.       \label{target_selection_now}}
\vspace{-0.1cm}
\end{center}
\end{figure}

In the target-limited regime, there are precious few bright targets to observe, so they are observed with all possible instruments (Figure~\ref{target_selection_now}).  This describes the current state of transiting planet atmosphere observations: the vast majority of atmospheric observations have focused on the few brightest targets \citep[$\sim 50$ planets have had their thermal emission measured with \emph{Spitzer}, but these data only serve to make estimates on the planets' dayside emitting temperature;][]{Cowan_2011b, Hansen_2014}.  

In the time-limited regime, on the other hand, there is an embarrassment of riches.  The challenge is to determine which targets are most interesting for follow-up work. There are two reasons to believe that we will soon be entering the time-limited regime for hot Jupiter and warm sub-Neptune characterization.  Firstly, the large collecting area of the JWST ensures that current borderline targets will yield high-precision observations.  Secondly and more importantly, TESS will provide thousands of targets that are as bright as our current darling planets.

Planets that are harder to detect, because of their small size and/or long orbit, will remain in the target-limited regime even with the advent of TESS.  It is a safe bet that JWST will study all the bright exemplars. Moreover, the small size and cool temperatures of such rare worlds make it likely that JWST would be the only space observatory capable of characterizing these worlds, which will only strengthen the case for studying them with the flagship telescope (ELTs may perform transit spectroscopy on such cool planets, but will not be able to detect their thermal radiation). 

\subsection{A Straw-Man Schedule for JWST}
Temperate terrestrial planets discovered by TESS will probably not be confirmed until JWST's second year of observations.  It therefore seems likely that the first two years of transiting exoplanet time will be spent characterizing the atmospheres of currently known (or soon to be discovered) hot Jupiters and warm sub-Neptunes.  \emph{In principle, this could yield credible atmospheric abundances and vertical temperature structure for about 50 short-period giants.}  Our community should be wary of observing more planets at the price of less complete spectral coverage and poor signal-to-noise; this is one of the important lessons learned from \emph{Spitzer}.   As observers and instrument specialists learn about the telescope's abilities and limitations, it may also be possible to perform spectrally resolved phase mapping of a half-dozen choice planets.

Once there are confirmed temperate terrestrial planets transiting bright M-Dwarfs in the JWST continuous viewing zone ($\sim$2020), a large fraction of the telescope's transiting planet time could and should be devoted to characterizing their atmospheres.  \emph{These measurements will be time-intensive and unlikely to conclusively establish or refute the habitability of these worlds, but would lay the ground-work for future flagship missions.}                                      

\section{Maximizing Exoplanet Science}
The next decade of transiting planet characterization is bound to be more exciting than the last.  The challenge will be to ensure that the science yield keeps pace with the excitement.

\subsection{The Case for a Dedicated Exoplanet Atmosphere Survey Telescope}
\textbf{JWST will have insufficient time to study the hundreds--thousands of hot Jupiters and warm sub-Neptunes discovered by TESS, and these observations will be heterogeneous both in terms of spectral coverage, observing mode, and data reduction. The best way to capitalize on the TESS discoveries is with a dedicated exoplanet atmosphere survey.  A meter-class space telescope with an optical---NIR spectrograph could perform a survey of the brightest non-terrestrial TESS planets in 1--2 years.}

There is a compelling portfolio of transiting exoplanet science that can be achieved with JWST, but two important caveats: (1) if the characterization of temperate terrestrial planets is feasible, it will likely take up much of the exoplanet time allotment in the 2020's, and (2) there is currently no intelligible plan for choosing which of the thousands of bright TESS giant planets to characterize.\footnote{It is also possible that the very nearest TESS targets will be too bright for JWST to observe, or at least to observe efficiently.  For example, various NIRSpec and NIRISS modes have saturation limits of $J$=7--5  \citep{Beichman_2014}.} We discuss each of these problems in turn below.

If one accepts the hypothesis that target-limited planets will be observed at any cost, then observations of potentially habitable worlds could take most of the $\sim640$~days of transiting planet time we have hypothesized the exoplanet community might obtain over a ten year mission. In particular, all estimates indicate that even a basic characterization of a temperate terrestrial planet will take of order 100~days. One could easily spend the entirety of JWST transiting planet time following up the 4 most habitable TESS planets. The only scenario in which JWST does \emph{not} spend hundreds of days staring at cool rocky planets is if its instruments suffer from such severe systematics that such observations are deemed a waste of time; this scenario seems unlikely.

There is only enough transiting exoplanet time in a 5~year JWST mission to fully characterize 25 short-period giant planets (full-orbit observations with full spectral coverage).  It is not obvious how to partition this time between hot Jupiters and warm sub-Neptunes, between circular and eccentric orbits, or between host star spectral type in order to maximize the scientific returns.  If \emph{Spitzer} is any indication, the temptation will be to use the majority of the telescope time on the smaller, cooler targets.  The question of target selection is further complicated by the existence of certain rare classes of objects.  For example, a dozen hot Jupiters have precise \emph{Kepler} eclipse measurements \citep{Heng_2013}, but only one has an appreciable albedo indicative of clouds \citep[Kepler-7b;][]{Demory_2013}. Likewise, many transit spectra are washed out by atmospheric hazes \citep{2014Natur.505...66K, 2014Natur.505...69K}, but some are haze-free \citep{Fraine_2014}; it would be wasteful to obtain countless featureless spectra with the premier space observatory. The scientific value of JWST observations would be maximized by focusing on archetypal planets rather than simply going after the brightest targets. 

\begin{figure}[!htb]
\begin{center}
\vspace{-0.1cm}
      \includegraphics[width=70mm]{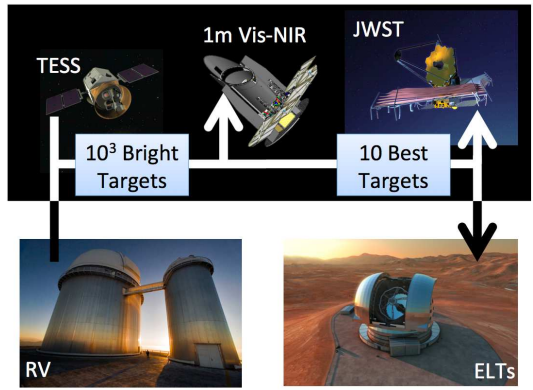}  
    \caption{\emph{Time-Limited Regime:} A combination of ground-based RV monitoring and space-based transit surveys (primarily TESS, but also Cheops, and eventually PLATO), will deliver thousands of transiting planets comparably bright to the current best targets. There will be insufficient time to perform atmospheric follow-up observations with JWST and ELTs.  However, these targets will be sufficiently bright that a modest telescope could obtain compelling atmospheric constraints. Moreover, the duration of transit and eclipse observations are limited by planetary dynamics, so a serial transit survey (one system at a time) could observe $10^3$ planets per year.  This motivates a dedicated exoplanet atmosphere survey telescope to obtain uniform observations of the brightest transiting planets, and to determine the few non-terrestrial planets most worthy of follow-up observations. Since the targets would be bright and have large scale heights, such a mission could be relatively modest: $\sim$1~m diameter primary mirror coupled to a optical--NIR spectrograph. \label{target_selection_later}}
\vspace{-0.1cm}
\end{center}
\end{figure}

\emph{In short, there is a clear need for a dedicated telescope to perform a uniform survey of hundreds of bright transiting giant planets in order to allow JWST to focus on (1) the smaller, cooler planets that it is uniquely able to characterize, and (2) the dozen archetypal giant planets worthy of detailed study (Figure~\ref{target_selection_later}).}

Based on \S\ref{benchmarks}, \emph{a $\sim1$~m space telescope equipped with an optical---NIR spectrograph should be able to measure---in a single occultation---the molecular absorption features of bright ($J<11$) TESS sub-Neptunes, and to detect the reflected light and thermal emission of the very brightest ($J<9$) hot Jupiters.} We expect such a mission to cost \$500M, to within a factor of 2. Alternatively, a scaled-down version of this survey could be performed using a probe-class ($\lesssim$ \$1B) direct-imaging mission without its coronagraph/starshade.

There is a trade-off between the capabilities of such a survey instrument and its timeliness.  In order for it to act as a triage for JWST sources, the mission would need to fly as soon as possible and would nessecarily be relatively modest.  A more ambitious mission, on the other hand, would likely fly towards the twilight years of the JWST mission, but would itself be able to provide detailed characterization of hot Jupiters and warm sub-Neptunes. 

An easier question to address is how many transiting planets such a survey telescope could/should observe.  First of all, the shortest segment of time that makes sense for observing a transiting planet is 3$\times$ the transit duration, or 6 hours, for typical short-period planet with 2-hour transits.  Even telescopes with very large collecting areas must observe planets for at least this amount of time. Secondly, bright transiting targets are few and far between, dictating that an atmospheric characterization mission will necessarily study them serially rather than in parallel.  As a result, a transit characterization mission can observe up to $365\times4\approx1000$~planets per year, regardless of telescope diameter. By lucky coincidence, this number is well-matched to the number of bright ($J<11$) transiting planets that TESS is expected to find. 

For the shortest-period planets, eclipse and phase measurements could additionally constrain reflected light and thermal emission. Depending on the telescope diameter and the brightness of target stars, some systems might benefit from multiple observations to beat down the photon noise. Repeat measurements would also enable robust inference of uncertainties, although a dedicated mission may be so well calibrated that uncertainties can be adequately estimated based on a single event.  These sorts of considerations ---not to mention scheduling constraints--- would realistically limit a dedicated mission to observing hundreds of transiting exoplanets in one year, roughly lining up with the expected number of TESS candidates with $J<10$, not to mention transiting planets discovered by the K2 \citep{Barclay_2014}, CHEOPS \citep{Broeg_2013}, and PLATO \citep{Rauer_2014}, or the countless ground-based wide-field transit searches.  

\subsection{The Case for M-Dwarfs}
\textbf{JWST will be the best observatory for constraining the atmospheres of temperate terrestrial planets orbiting nearby M-Dwarfs, but will only be able to obtain very basic atmospheric constraints: establishing the presence of an atmosphere, measuring its emitting temperature, detecting greenhouse gases, and---if orbital phase monitoring is possible---constraining the planet's rotational state, surface pressure, and the presence of clouds. These constraints will be tantalizing, but in order to properly characterize these worlds, future flagship missions will have to be designed with M-Dwarf planets in mind.}

It is generally accepted that the next big step in exoplanet atmospheric characterization after JWST will be a direct-imaging mission to search for and characterize the atmospheres of Earth-like planets orbiting Sun-like stars \citep[e.g., ATLAST, LUVOIR;][]{Kouveliotou_2014}.  In order to minimize risk and maximize scientific return, such a flagship mission must also be able to directly image habitable planets orbiting nearby M-Dwarfs. 

The frequency of temperate terrestrial planets is already known to be high for M-Dwarfs \citep{Dressing_2013, Morton_2013, Dressing_2015}, and the TESS mission is predicted to find a few nearby transiting examples. On the other hand, not a single Earth analog has yet been detected in orbit around a G-Dwarf, and the frequency of such planets therefore requires extrapolation to unexplored regions of parameter space \citep{Petigura_2013, Foreman-Mackey_2014, Farr_2015}. It might be possible to interpolate over (rather than extrapolate to) the habitable zone of Sun-like stars by combining the microlensing statistics of WFIRST with the transiting statistics of \emph{Kepler} in order to obtain a more robust frequency by the late 2020's.  In order to find \emph{nearby} Earth twins, we will need 10~cm/s radial velocity surveys, a dedicated astrometry mission, or an ambitious direct-imaging survey.  It therefore appears likely that a next-generation direct-imaging flagship mission will be designed before we know the frequency of terrestrial planets in the habitable zones of Sun-like stars, let alone the location of the nearest examples.  In order to minimize the risk of such a blind search, it would be prudent to design the telescope such that it could characterize temperate planets orbiting \emph{low}-mass stars. Such planets will be discovered in the next decade, and their atmospheric characterization will have begun in the JWST era.

\section{Summary}
\textbf{JWST will completely characterize a few dozen hot Jupiters and sub-Neptunes, and obtain rudimentary constraints on the atmospheres of Earth-like worlds.  This leads us to two conclusions: (1) a dedicated mission is needed to perform an atmospheric survey of TESS planets. Most of these planets are hotter and larger than Earth, with atmospheres made of lighter molecules, and therefore relatively easy to observe: a 1~m space telescope equipped with an optical---NIR spectrograph could perform a uniform atmospheric survey for hundreds of the brightest TESS planets; (2) the detailed characterization of temperate terrestrial planets orbiting M-Dwarfs should be an explicit goal of future flagship missions.} 

\acknowledgements{NBC acknowledges the generous hospitality of l'Institut de Plan\'etologie et d'Astrophysique de Grenoble, where he wrote much of this report.}

\begin{deluxetable}{lcccccccc}
\rotate
\tabletypesize{\scriptsize}
\tablecaption{White Light Photon-Counting Precision of 1~hr Integrations on a Target at 20~pc$^a$ \label{precisions}}
\tablewidth{0pt}
\tablehead{
&\colhead{JWST/NIRSpec} & \colhead{HST/WFC3}& \colhead{JWST/NIRSpec} & \colhead{Spitzer/IRAC} & \colhead{JWST/NIRCam} & \colhead{JWST/MIRI} & \colhead{Spitzer/MIPS}&\colhead{JWST/MIRI}\\
&\colhead{0.6--1.0~$\mu$m} & \colhead{G141} & \colhead{1.0--4.0~$\mu$m} & \colhead{ch1} & \colhead{F356W} & \colhead{F1000W} & \colhead{24~$\mu$m}&\colhead{F2550W}
}
\startdata
$D$& 6.5~m& 2.4~m& 6.5~m& 0.85~m&6.5~m& 6.5~m&0.85~m&6.5~m\\
$\lambda_1$& 0.6~$\mu$m& 1.1~$\mu$m& 1.0~$\mu$m& 3.225~$\mu$m& 3.12~$\mu$m& 9~$\mu$m& 21.65~$\mu$m &23.5~$\mu$m\\ 
$\lambda_2$& 1.0~$\mu$m& 1.7~$\mu$m& 4.0~$\mu$m& 3.975~$\mu$m& 4.01~$\mu$m& 11~$\mu$m& 26.35~$\mu$m& 27.5~$\mu$m\\
$\tau$& 0.30 & 0.40 & 0.40 & 0.40 & 0.40 & 0.36 & 0.45 & 0.18\\
& & & & & & &\\
5000~K & $8.8\times10^{-7}$ & $2.2\times10^{-6}$& $5.4\times10^{-7}$& $1.6\times10^{-5}$&$1.8\times10^{-6}$&$5.3\times10^{-6}$&$8.4\times10^{-5}$&$2.1\times10^{-5}$\\
3000~K & $1.1\times10^{-5}$ & $1.8\times10^{-5}$& $3.8\times10^{-6}$ &$8.9\times10^{-5}$&$1.0\times10^{-5}$& $2.7\times10^{-5}$& $4.2\times10^{-4}$& $1.0\times10^{-4}$\\
\enddata
\tablecomments{$^a$These white light precisions only directly apply to measurements of reflected light, thermal emission, or broadband spectroscopy.  The precision for bona fide spectroscopy is less impressive.  For example, if one wanted to obtain a 1.0--4.0~$\mu$m spectrum with 0.1$\mu$m resolution using JWST/NIRSpec ($R\approx20$), then the expected precision would be $\sqrt{30}\approx 5.5$ times worse than the values shown. These photon-counting estimates include imperfect instrument throughput, but not read-noise, dark current, sky background, nor any detector or astrophysical systematics.  The precisions for HST and Spitzer are computed the same way and are close to current empirical precisions (\S\ref{comparison}). Nonetheless, it is possible than MIRI observations will be limited by warm telescope background rather than photon counting, since JWST will be passively cooled.}
\end{deluxetable}

\end{document}